\documentstyle[12pt]{article}
 at 14.4truept at 12.0truept
\title{\large\bf 
Self-Dual ${\mathcal{N}}$=8 Supergravity\\[8pt] as a Closed 
String Field Theory in Twistor Space}
\author{
{\large Satchidananda  Naik}
\\
  Harish-chandra Research Institute \\
 Chhatnag Road, Jhusi  \\
Allahabad-211 019, INDIA\\}

\begin{document}
\maketitle

\hspace*{\fill}

\hspace*{\fill}
\newcommand{\bee}{\begin{equation}}
\newcommand{\nn}{\nonumber}
\newcommand{\ee}{\end{equation}}
\newcommand{\ba}{\begin{array}}
\newcommand{\ea}{\end{array}}
\newcommand{\bea}{\begin{eqnarray}}
\newcommand{\eea}{\end{eqnarray}}
\newcommand{\ki}{\chi}
\newcommand{\eps}{\epsilon}
\newcommand{\pa}{\partial}
\newcommand{\lb}{\lbrack}
\newcommand{\Se}{S_{\rm eff}}
\newcommand{\rb}{\rbrack}
\newcommand{\de}{\delta}
\newcommand{\th}{\theta}
\newcommand{\rh}{\rho}
\newcommand{\ka}{\kappa}
\newcommand{\al}{\alpha}
\newcommand{\bt}{\beta}
\newcommand{\si}{\sigma}
\newcommand{\basi}{\bar \psi}
\newcommand{\bsi}{\Sigma}
\newcommand{\vp}{\varphi}
\newcommand{\gm}{\gamma}
\newcommand{\gb}{\Gamma}
\newcommand{\om}{\omega}
\newcommand{\et}{\eta}
\newcommand{\qab}{{{\sum}_{a\neq b}}{{q_a q_b}\over{R_{ab}}}}
\newcommand{\omb}{\Omega}
\newcommand{\pr}{\prime}
\newcommand{\ra}{\rightarrow}
\newcommand{\nb}{\nabla}
\newcommand{\MSb}{{\overline {\rm MS}}}
\newcommand{\lnh}{\ln(h^2/\Lambda^2)}
\newcommand{\cz}{{\cal Z}}
\newcommand{\h}{{1\over2}}
\newcommand{\lm}{\lambda}
\newcommand{\Lm}{\Lambda}
\def\lmb{{\bar{\lambda}}}
\newcommand{\inft}{\infty}
\newcommand{\bpa}{\bar {\partial}}  
\newcommand{\CN}{\mathcal{N}}
\newcommand{\CA}{\mathcal{A}}
\newcommand{\hth}{\hat {\theta}}
\newcommand{\hp}{\hat p}
\newcommand{\hb}{\hat b}
\newcommand{\hc}{\hat c}
\newcommand{\hbt}{\hat {\beta}}
\newcommand{\hgm}{\hat {\gamma}}
\newcommand{\hvp}{\hat {\varphi}}
\newcommand{\hlm}{\hat {\lambda}}    
\newcommand{\hv}{\hat v} 
\newcommand{\hq}{\hat q} 
\newcommand{\abs}[1]{\left\vert #1\right\vert}
\def\dvec#1{\buildrel \leftrightarrow \over #1}
\def\sfrac#1#2{{\textstyle\frac#1#2}}
\def\dt#1{{\buildrel {\hbox{\bf .}} \over {#1}}}  
\def\ddt#1{{\buildrel {\hbox{\bf ..}}\over {#1}}}
\def\ald{\dt{\alpha}}
\def\btd{\dt{\beta}}
\def\gmd{\dt{\gamma}}
\def\ded{\dt{\delta}}
\def\aldd{\ddt{\alpha}}
\def\btdd{\ddt{\beta}}
\def\dedd{\ddt{\delta}}
\def\pd{{\dt{+}}}
\def\md{{\dt{-}}}
\def\dz{{\dt{0}}}
\def\ddz{{\ddt{0}}}
\def\ddo{{\ddt{1}}}
\newcommand{\Lra}{\Longleftrightarrow}
\newcommand{\abschnitt}[1]{\par \noindent {\large {\bf {#1}}} \par}
\newcommand{\subabschnitt}[1]{ \noindent
                                          {\normalsize {\it {#1}}} \par}
%
%
%
%
\newcommand\dsl{\,\raise.15ex\hbox{/}\mkern-13.5mu D}
\newcommand\delsl{\raise.15ex\hbox{/}\kern-.57em\partial}
\newcommand\Ksl{\hbox{/\kern-.6000em\rm K}}
\newcommand\Asl{\hbox{/\kern-.6500em \rm A}}
\newcommand\Dsl{\hbox{/\kern-.6000em\rm D}} 
\newcommand\Qsl{\hbox{/\kern-.6000em\rm Q}}
\newcommand\gradsl{\hbox{/\kern-.6500em$\nabla$}}
\newpage
\begin{abstract} \normalsize
 A closed string field theory action is formulated for the
$\CN$ = 8 self-dual supergravity which is off-shell and
Lorentz covariant. The bosonic truncation in the quantum
field theory limit gives the Plebanski action in the
super space.
\end{abstract}

\vskip10.0cm
\newpage
\pagestyle{plain}
\subabschnitt{\bf 1. Introduction:}
The conventional quantum field theory calculation of gravity/supergravity
loops in four dimensions is maligned with ultraviolet divergences and 
hence it
is non-renormalizable \cite{THF}.This is easily realized by naive power
counting and the presence of higher derivative terms in the Lagrangian. On 
the contrary  it is claimed that $\CN$ = 4 super-Yang Mills theory is
a finite theory \cite{SYF}. By Kawai, Lewellen and Tye (KLT)rule the 
gravity tree amplitudes can be expressed as the Yang-Mills tree 
amplitudes \cite{KLT}. By the method of unitarity cutting rules it is
shown that $\CN$ = 8 supergravity has the same ultraviolet behaviour
as that of $\CN$ = 4 super-Yang Mills theory and hence it is
claimed to be finite 
\cite{BBB}. This fact renews  ample motivation to study off-shell
behaviour of $\CN$ = 8 supergravity in the form of closed string field 
theory. The usual  $\CN$ = 8 supergravity action is obtained from 
compactifying
type II string theory on a six torous or $d = 11$ and $\CN$ =1 
supergravity M-theory action. On the contrary we study here the self-dual
supergravity as a $\CN$ =2  closed string field theory in the
Attiyah-Ward (AW) space-time. The reason being it is simple and in the 
linearized level it is same as full  non-selfdual theory. The action has 
the same field 
content where fields with negative helicities are treated  as Lagrangian 
multiplier to the gauge covariant kinetic terms of the positive helicity 
states. In the pioneering work  of Witten on twistor string theory 
\cite{WITT},
 all the  maximally helicity violating (MHV) amplitudes of
$\CN$ = 4 super-yang Mills theory are reproduced as {\bf D}-instanton 
corrections to the topological {\bf B}-type string 
where a Calabi-Yau supermanifold
 ${\mathcal C}P^{3|4}$ space emerges naturally  as the target space of the 
super-twistor string.
The Lagrangian for this {\bf B}-model is the Open String Field theory
Lagrangian of super-holomorphic Chern-Simons theory, which happened to
be anti-self-dual Super Yang-Mills theory. Here the {\bf D}-instanton 
correlation function plays the vital role in producing the full non-
self- dual  super Yang-Mills amplitudes. However Berkovits
\cite{BERV1} gives an alternative formulation for the same MHV amplitudes
as the conventional open string amplitudes of the super twistor string.
There is a natural generalization of this twistor string proposal of   
$\CN$ = 4 Super-Yang-Mills theory  to that of
 $\CN$ = 4 conformal supergravity \cite{BEWT}. The conformal 
supergravity is very nice in the classical level however the quantization
is not very physically interesting due to the presence of 
higher derivative terms even in the linearized approximation. Due to this 
Hilbert space is not positive definite and consequently  S-matrix violates 
Unitarity. So a covariant string  formulation of  $\CN$ = 8 Einstein 
Supergravity is more desirable. There are attempts to formulate a super 
twistor string theory for $\CN$ = 8 supergravity \cite{AHM}.
On the contrary we start from a $\CN$ = 2 critical closed string theory
and write a covariant  string field theory lagrangian for self-dual
$\CN$ = 8 supergravity. Recently a Lorentz covariant Open string field 
theory Lagrangian  for $\CN$ = 4 self-dual Super-Yang-Mills
theory has been proposed by Lechtenfeld and Popov \cite{LP}.
In the same spirit we are trying to formulate a closed string field
theory for the $\CN$ =8 supergravity. However the problems and 
intricacies are infinitely bigger for this case than the Super-Yang
Mills. First of all to show the local gauge invariance of the 
super-space one needs to generate a target super-space
from the world sheet variables. Generating super spcae from
the Neveue-Schwarz variables is extremely complicated may be
possible. Thus we followed Siegel's \cite{SIEG} approach of
putting non-holomorphic scalar grassman variables to formulate
the super-space.

The organization of the paper is as follows. We give adequate review
of the $\CN$ =2 critical string theory and basic tools needed for the
further construction. First part of section 3. has basic review  
of Berkovit's formulation of Closed string and also open string
for the Self-dual Yang Mills. In the later part of section 3.
 we derive our results. 

\subabschnitt{\bf 2. The Preliminaries:}
 The self-dual  Yang-Mills and self-dual Gravity theories in $2+2$
dimensions are described by $\CN$ = 2 critical open and closed strings 
respectively \cite{N2S}. The perturbative spectrum of $\CN$ = 2 string
has only $p^2~=~0$ as the only ground state. For the four dimensional 
Euclidean space as the target space  there will be each $p_i$ zero so that
there will be no dynamics and hence this theory is not at all interesting.
For the  non-trivial  dynamics of the physical ground state, the target 
space need to have an $SO(2,2)$ symmetry so that  momenta can be complex. 
Also only non-vanishing S-matrix amplitude is the three point function
and all higher n-point functions are zero due to the kinematics of $(2,2)$
space. So this massless scalar of the ground state of Open/closed string 
describes the dynamics 
of self-dual Yang-Mills/ Gravity which is topological in 
nature. Before proceeding further it is worthwhile  to address
the space-time supersymmetry of the 
$\CN$ = 2 fermionic string.
The three point function is only invariant under the subgroup
$U(1,1)$ of the total Lorentz group $SO(2,2)$ of the target space.
This makes the  the space-time supersymmetry of $\CN$ = 2 critical
string theory obscure. However the ground states of the fermionic
string with $\CN$ = 4 world-sheet supersymmetry has larger symmetry
and is covariant under $SU(~1,~1)~\times ~SU(~1,~1)$. Since the highest 
weight 
states of $\CN$ = 4 Superconformal algebra are the same as the physical
states of $\CN$ = 2 algebra, it has been argued that
both the theories are one and the same
\cite{SIG}. The four world-sheet
fermions with the combinations of both periodic and
 antiperiodic boundary conditions 
have sixteen bosonic and sixteen fermionic states as the ground states 
for the closed string and the spectrum is clearly supersymmetric. 
To  realize  explicit 
Lorentz covariant one needs these  extra missing symmetry
 $SO(~2,~2)/SU(~1,~1)$ which happened to be $SL(~2,~R)$ as twistor 
variables.
 The Brink and Schwarz~\cite{BS} action for $\CN$ = 2 string is:
 \bea
  S &=&\ \int d^2\si~ \sqrt{g}\,\Bigl\{
          \sfrac12 g^{ab}\pa_a x^{i} \pa_b {\bar x}^{\bar i} 
         +\sfrac{i}2 \basi^{\bar i}{\gm}^{a} \dvec{D}_a {\psi}^{i}
         +A_a \basi^{\bar i}{\gm}^{a}{\psi}^{i}  \nonumber\\[.7ex]
      && +\,(\pa_a {\bar x}^{\bar i}~ +~ \basi^{\bar i}{\chi}^{+}_a)
          {\chi}^-_b {\gm}^a{\gm}^b {\psi}^{i}
         + h.c.  \Bigr\}\,{\et}_{{\bar i} i},
\eea
where $g^{ab}$ and $A_a$ with $a{=}0,1$, are the (real) worldsheet
metric and $U(1)$ gauge connection, respectively and $D_a$ is the 
gravitational covariant derivative for the spinor with 
 spin connection in two dimension and $\gm_a$ are the worldsheet
gamma matrices.
 Here $x^i$ and ${\psi}^i$ are complex
valued in the target space with ${\et}_{{\bar i} i}$ metric which has a 
complex structure. In the $\CN$ = 2   superconformal gauge one has both
 left and right hand sectors of superconformal constraint algebras 
denoted as $T$, $G^{+}$ , $ G^{-}$
and $J$ and ${\hat T}$, ${\hat G^{+}}$ , ${\hat {G^{-}}}$ and 
${\hat J}$
respectively. In both the sectors one has usual (b,c) ghosts,a pair
of 
($\bt_{\pm}$, $\gm_{\pm}$ )ghosts and a pair of (u,v) ghosts for 
$U(1)$ gauge
fixing. The cancelation of anomaly dictates the central charge of the
matter sector to be 6.
 It is more convenient for our purpose to use 
$SL(2, R)\times SL(2, R)$ basis instead of complex coordinate basis
for the target space. In $SL(2, R)\times SL(2, R)'$ spinor notation
\bee
x^{\al \ald}~ = ~ \si_\mu^{\al \ald} x^\mu ~ =~
\left(\begin{array}{cc}
x^4{+}x^2 & x^1{-}x^3 \\ x^1{+}x^3& x^4{-}x^2
\end{array}\right) \quad,\qquad
\al\in\{0,1\} \; , \quad \ald\in\{0,1\} \;,
\ee
where $\si_\mu$ describes  the chiral $\gm$ matrices. However in these 
basis the associated constraints in the left moving sectors are
\bea
T\ =\ \pa_z x^{\al \btd}\pa_z x_{\al \btd}
     +\psi^{\ald\btdd}\pa_z\psi_{\ald\btdd}\quad,
\eea
\bee
G^{0\ddo}\ =\ \psi^{\ald\ddo} \pa_z x^0\!_\ald \quad, \qquad
G^{1\ddz}\ =\ \psi^{\ald\ddz} \pa_z x^1\!_\ald \quad,
\ee
\bee
J^{\ddz \ddo} ~~= ~~ \psi^{\ald\ddz} \psi_\ald^{\ \ddo}.
\ee
There are similar constraints for the right moving sectors as
well.
Here $\al$, $\bt$, $\ald$, $\btd$ are space-time spinor 
indices and $\aldd$, $\btdd$ are world-sheet $SL(2, R)'$ internal
indices for the $\CN$=2 R-symmetry.
The operator product expansions for $X$ and $\psi$ are
\bee
x^{\al \ald}(z,\bar z)\ x^{\bt \btd}(w,\bar w)\ \sim
{\eps}^{\al\bt}{\eps}^{\ald \btd}\,\ln |z-w|^2 \quad,
\psi^{\ald \btdd}(z)\psi^{\gmd \dedd} (w)\sim
\frac{2{\eps}^{\ald \gmd}~{\eps}^{\btdd \dedd}}{z-w}\,.
\ee
In case we bosonize the fermions, the
$U(1)$ current $J^{\ddz \ddo}$ becomes $\pa H$. It is noticed
 by Berkovits and Vafa \cite{BV} that commutation relations among
${\exp} ~ ( H)$, ${\exp}~ (-  H)$ and $\pa H$  form an
$SU(2)$ or $SU(1,1)$ algebra. Incorprating these as new currents,
$\CN$ = 2 superconformal 
algebra
is extended to small $\CN$ = 4 superconformal algebra as
\bee
G_{\al}^{\ \ \btdd} = \psi^{\gmd\btdd}\,\pa_z x_{\al\gmd}\ ,\qquad
J^{\aldd\btdd} = \psi^{\gmd\aldd}\,\psi_\gmd^{\ \ \btdd} .
\ee
There is an elegant way to make the theory with vanishing central
charge without ghosts by twisting the energy-momentum tensor
$T\to T':= T + \frac{1}{2}\pa J^{\ddz \ddo}$. By this the conformal
weight of the  spinor  $\psi^{\ald \ddz}$ has become zero and that of
$\psi^{\ald \ddo}$ has become one. Both the fermions are now integer 
moded and also fermion of zero weight has zero modes on the sphere.
Due to the twisting the curvature singularities of the
world sheet are taken care of by
these twisted fermions rather than  the ghosts. In a sense the $U(1)$
anomaly plays the role of the ghost anomaly and $\psi$'s play the 
role 
of the $(b,c)$ ghosts. Consequently $G_{\al}^{\ \ \ddz}$ has spin
one  with conformal weight one and $G_{\al}^{\ \ \ddo}$ has spin two
with conformal weight two. Hence $G_{\al}^{\ \ \ddz}$ play the role
of BRST- type of  currents. Similarly $J^{\ddz \ddz}$ and 
 $J^{\ddo \ddo}$ have conformal weight zero and two respectively.
This  theory has now $\CN$ = 4 worldsheet super symmetry and also
has $SL(2,R)\times SL(2,R)'$ space-time symmetry.
In order to formulate $\CN$ = 8 supergravity we keep
one of the $SL(2,R)$ symmetries rigid and another
to extend it to ${\mathcal{OSP}}(8\mid 2)$ and consequently gauge
it.This needs eight fermionic variables to form the 
superspace. Infact the ground state closed string
spectrum  has eight such states. However it is quite
obscure to use these as super cordinates. Following
Siegel \cite{SIEG} we introduce eight fermionic 
coordinates ${\th}^{\al i}$ as super partner of
$x^{\al \ald}$, which are worldsheet scalars with
non-holomorphic corelation function. Inorder to
maintain the conformal invariance a set of bosonic
variables $(b^i, c^i)$ are introduced. The operator
product expansions of $\th$'s and for $(b^i, c^i)$ are
\bee
{\th}^{\al i}(z,\bar z)\,{\th}^{\bt j}(w,\bar w)\ \sim
{\eps}^{\al\bt}{\de}^{i j}\,\ln |z-w|^2 \quad,
b^{i}(z)\,c^{j} (w)\sim
\frac{{\de}^{i j}}{z-w}\,.
\ee
The associated stress tensor is 
\bea
T\ =\ \pa_z x^{\al \btd}\pa_z x_{\al \btd}
     +\psi^{\ald\btdd}\pa_z\psi_{\ald\btdd}
     +\h\pa_z {\th}^{\al i}\pa_z{\th}^i_{\al}
     + b^{i}\pa_z c^i
\quad,
\eea
and also similarly its super partners
\bee
G^{\al\ddo}\ =\ \psi^{\ald\ddo} \pa_z x^{\al}\!_\ald 
+ c^i\pa_z {\th}^{\al i}
\quad, \qquad
G^{\al \ddz}\ =\ \psi^{\ald\ddz} \pa_z x^{\al}\!_\ald 
+ c^i\pa_z {\th}^{\al i} 
\quad,
\ee
\bee
J^{\ddz \ddo} ~~= ~~ \psi^{\ald\ddz}
 \psi_\ald^{\ \ddo} + c^i~ b^i.
\ee
However the OPE of these $J^{\ddz \ddo}$'s are
singular as ${\frac{2 + \CN}{z^2}}$ indicating a
$U(1)$ anomaly with charge $-(2 +\CN)$ which are
 2 due to $\psi_\ald^{\ \ddo}$ and $\CN$ due to
$(b^i~c^i)$ systems respecively.
In the sequel we will show the
 proper insertions of these $\psi$'s and 
$c^i$ 's  for taking the vacuum expectation
 values of the vertex operators.( For the closed string
all the right moving anti-holomorphic fields are
described as hatted variales.)

\subabschnitt{\bf 3. Covariant Closed  String Field theory:}\noindent  
The most conventional approach to covariant String field theory is 
Witten's Open String Field theory (OSFT) \cite{WOSFT}, where strings 
join together at their mid point. The gauge invariance of the OSFT 
demands only a  cubic term in its interaction.  This procedure of
joining and splitting of  interacting  strings 
 maintains the length of the string intact. If one extends this 
approach to closed string and trys to maintain the length of the
intercting string fixed then the required prescription would be
to overlap one of the closed strings with the other half of another 
closed string by that the resulting  closed string will have
the same length as that of  the overlapping strings. The gauge 
invariance of this type of gluing procedure demands the action to be
non-polynomial in string field \cite{ZEW}. There is another school
of thought for covariant string field theory pursued by 
Hata et al. \cite{HIKKO}, who generalized the light-cone string field
theory to covariant one by BRST procedure. In this approach the open 
strings join at their end points and closed string join together at
one point only. The gauge invariance of the string fields requires
that closed string should have cubic interaction where as open string
should have non-polynomial interaction term.
 However by this the string length cannot be held 
fixed hence behave as a free parameter in the theory. Although this
parameter is easily gauged away in the on-shell limit, the ambiguity
remains for the off-shell theory.
 Despite this ambiguity we follow
this formulation for the simplicity due to the cubic 
interaction. 

\subabschnitt{\bf 3.1 Preliminaries of closed string field theory 
action:}\noindent The basic string field $\Phi$ is a 
functional of the worldsheet string coordinates 
$\{{x^{\al \btd}},\psi^{\ald\btdd}
{\hat{\psi}}^{\ald\btdd},{\th}^{\al i},
b^i,{\hat b}^i, c^i,{\hat c}^i \}$.
The basic ingredints needed for the gluing 
and splitting of the closed strings are prescription
dependent and these are presented in te appendix.   
For illustration we present here the Berkovits
$\CN$ = 2 closed string theory \cite{BERKF}.
For the gluing of string fields,
 a  curvature is created at the joining points
which is related to the $U(1)$ anomaly 
as explained in the appendix..
 The back-ground charge due to the $U(1)$
current which is equivalent to ghost current
 of $(b, c)$ system
is  $-\frac{D}{2}$ which happens 
to be -2 in this case. One needs
insertion of zero modes of $\psi^{\ald \ddz}$ 
to compensate this 
charge in each vertex operator. Taking this in to
account 
Berkovits  \cite{BERKF} has prposed a
closed string field theory action for $\CN$ = 2
 string field theory
 which kinetic term is given as:
\bee
S=\int < \Phi  (J^{\ddz\ddz}_0 + {\hat J}^{\ddz\ddz}_0)
(G^{0 \ddz}_0 +{\hat G}^{ 0 \ddz}_0)(G^{1 \ddz}_0 + 
{\hat G}^{1 \ddz}_0) 
\Phi>
\ee
 Here subscript zero denotes the zero modes of the operators.
More explicitly $G^{0 \ddz}_0 ~=~ \psi^{\ald \ddz}
\frac{\pa}{\pa x^0_{\ald}}$ and similarly ${\hat G}^{ 0 \ddz}_0~=
~{\hat\psi}^{\ald \ddz}\frac{\pa}{\pa x^0_{\ald}}$ and using the 
contraction $<\psi^{\ald \ddz}_0\psi^{\btd \ddz}_0>$ as $\eps^{\ald 
\btd}$ one gets the correct kinetic term as
 \bee
   K.E ~=~ \eps^{\al \bt}\eps^{\ald 
\btd} \frac{\pa}{\pa x_{\al \ald}}\Phi
\frac{\pa}{\pa x_{\bt \btd}}\Phi.
\ee
The gluing of three string  $A$ , $B$ and $C$ satisfy $(A(BC)) = 
((BA)C) +((AC)B)$.
 The interaction term for covariant closed string is given 
as 
\bee
S_{int} = \int \Big[ (G^{1 \ddz}_0 -{\hat G}^{ 1 
\ddz}_0)\Phi,(G^{0 
\ddz}_0 
+{\hat G}^{ 0 \ddz}_0)(G^{1 \ddz}_0 -{\hat G}^{ 1 \ddz}_0)\Phi 
\Big]
(G^{0 \ddz}_0 +{\hat G}^{ 0 \ddz}_0)\Phi
\ee
 where $[,]$ is anti-symmetric product. After a lengthy calculation
taking in to account all sorts of contractions and permutations
we arrive at the required result which gives the Plebanshki action
as:
\bee
S_{int} = \int \eps^{\al \bt}\eps^{\gm \de}\Phi\frac{\pa}{ 
{\pa}{x_{\al \ald}}}\frac{\pa}{{\pa}{x_{\gm}^{\ald}}}\Phi 
\frac{\pa}{\pa {x_{\bt\btd}}}
\frac{\pa}{\pa {x_{\de}^{\btd}}}\Phi
\ee
where $\Phi$ is the pre-potential.

\subabschnitt{\bf 3.2 Berkovits-Siegel covariant string field theory 
for self-dual Yang-Mills:}\noindent
Berkovits and Siegel \cite{BS} have prposed a manifestly Lorentz 
covariant string field theory action for  self-dual Yang-Mills 
theory where string fields are the gauge connection
 ${\CA}_{\al}$ which makes the
BRST operator $Q_{\al}$ gauge invariant and the auxiliary field 
 $g_{\al \bt}$ which enforces the self-dual constraint.
 To illustrate
further -- the gauge covariant derivative 
$\nb_{\al}~=~G_{\al}^{\ddz}~+~{\CA}_{\al}$ and under the gauge
transformation 
\bee
\nb_{\al}' = \omb \nb_{\al}{\omb}^{-1},\quad \quad g_{\al \bt}' =\omb 
g_{\al \bt}{\omb}^{-1} + \nb^{\gm}K_{\gm \al \bt} 
\ee
where $\omb$ and $K_{\gm \al \bt}$ are gauge parameters of the
internal symmetry of the Yang-Mills gauge group such that the
Lorentz invariant string field theory action remains gauge invariant.
\bee
 S = \int <tr g_{\al \bt} f^{\al \bt}>.
\ee
 Here $\int $ is for integral over zero modes of  $x$ space and
$< ,>$ is the expectation value of string field operator products
where massive unphysical modes contribute in the off-shell,
and
\bee
f_{\al \bt} = \{\nb_{\al} ,\nb_{\bt}\} =        
G_{\al}^{\ddz}{\CA}_{\bt} +G_{\bt}^{\ddz}{\CA}_{\al}
+ \{{\CA}_{\al},{\CA}_{\bt}\}.
\ee
The $\{,\}$ denotes the symmetric product of the string fields
and each product is the Witten's OSFT  star product. Also
\bee
(G_{\al}^{\ddz}{\CA}_{\bt})(z) = 
\oint_z\,\frac{dw}{2\pi i}G_{\al}^{\ddz}(w){\CA}_{\bt} (z).
\ee
The equation of motion for the string fields from (13)gives
\bee
f_{\al \bt} = 0, \qquad G_{\al}^{\ddz}g^{\al \bt}~ 
+~ \lb{\CA}_{\al}, g^{\al \bt}\rb~= 0.
\ee
Berkovits and Siegel \cite{BS} have also generalized this action
for the $\CN$ = 4 Supersymmetric Yang-Mills theory. More 
recently Lechtenfeld and Popov \cite{LP}
 have given a covariant 
cubic string field theory action
for the self-dual $\CN$ = 4 Supersymmetric
 Yang-Mills theory in the supertwistor space.

\subabschnitt{\bf 3.3 Supertwistor space:}\noindent
As mentioned in the section 2. the perturbative
string spectrum has only manifest $SU(~1,~1)$
instead of space-time $SO(~2,~2)$ Lorentz symmetry.
This missing $SO(~2,~2)/SU(~1,~1)$
symmetry is treated as twistor transformation,
which will be introduced to rotate all the complex
structure of the $SO(2,2)$ and finally integration
over this parameter space will ensure the Lorentz
covariance. This geometric space
 ${\mathcal H}^2~=~SO(~2,~2)/SU(~1,~1)$ is a two
sheeted hyperboloid. For the Euclidean space $R^4$
the quotient space $SO(4)/U(2)$
parametrizes all the complex structure
happened to be the Rieman sphere $CP^1$.
Atiyah,Hitchin and Singer \cite{AHS} describe the
twistor space of $R^4$ as a cmplex
manifold ${\mathcal P}_E ~ = ~R^4\times CP^1$,which is
a rank 2 vector bundle over $CP^1$. similarly
 the twistor space of $R^{2,2}$
is a complex manifold ${\mathcal P}_H$ which is 
 $R^{2,2}\times{\mathcal H}^2$. 
This space is covered by two (acyclic
 i.e. in one of the patches ${\lm}^0$ is zero and in another
patch ${\lm}^1$ is zero ) coordinate
patches ${\cal U}_{\pm}$ with complex coordinates
$({\omega}^{\ald}_{\pm}, {\lambda}^{\al}_{\pm})$ on
each patches where
\bee
(\lm^\pm_\al)= 
\left(\begin{array}{cc} 1\\{\lm}_{\pm}\end{array}\right),
\quad(\lm^\al_\pm)= 
\left(\begin{array}{cc}{-{\lm}_{\pm}}\\1\end{array}\right),
\quad({\tilde{\lm}}^\al_\pm)=
\left(\begin{array}{cc}1\\-{\lmb}_{\pm}\end{array}\right)
\ee
with the normalization
\bee
\nu_+ = (1-,\lm_+\lmb_+)^{-1}, \quad\quad 
\nu_- = -(1-\,\lm_-\lmb_-)^{-1}
\ee
and ${\omega}^{\ald}_{\pm}~=~x^{\al \ald}{{\lambda}_{\al}}^{\pm}$.
On the overlap ${\cal U}_{-}\cap {\cal U}_{+}$ it is ${\lm_+}^{-}~=~\lm_-$.
Using the above normalization the inverse
tranformation of $x^{\al \ald}$ in 
terms of ${\omega}^{\ald}_{\pm}$ is easly found.
This shows the reparamtrization of the twistor space.
The ransformations of anti-holomorphic vector fields
and coordinate vector fields are
\bee
\frac{\pa}{\pa\bar{w}_\pm^\btd}=\nu_\pm
{\eps}_{\ald \btd}\lm_\pm^\al
\frac{\pa}{\pa x^{\al\ald}}~~,
\qquad
\frac{\pa}{\pa\bar{w}_\pm^{\dot{3}}}=
\frac{\pa}{\pa\lmb_\pm}- \,x^{1\ald}\lm^\al
\frac{\pa}{\pa x^{\al\ald}} \ .
\ee
Thus the anti-holomorphic vector fields are
\bee
\bar{v}^\pm_\ald ~ = ~{\lm}^\al_\pm 
\frac{\pa}{\pa x^{\al\ald}}~~, \qquad
\bar{v}^\pm_{\dot{3}} ~= ~\frac{\pa}{\pa\lmb_\pm}
\ee
Similarly one can construct holomorphic vector fields.
Let us extend this approach  to superspace geometry.
For the graded space $\{x^{\al \ald}, {\th}^{\al i}\}$, one has 
a graded twistor space \cite{MW,LP} as 
${\mathcal PS}_H$ which is
 $R^{(2,2)\mid \CN}\times{\mathcal H}^2$. The local coordinates on
patches ${\cal US}_{\pm}$ are $({\omega}^{A}_{\pm}, 
{\lambda}^{\al}_{\pm})$ where
${\omega}^{A}_{\pm}~=~ \lb x^{\al \ald}{{\lambda}_{\al}}^{\pm}
 ~~ {{\lambda}_{\al}}_{\pm}{\th}^{\al i}\rb$
The anti holomorphic vector fields are 
\bee
\bar{v}^\pm_\ald ~ = ~{\lm}^\al_\pm 
\frac{\pa}{\pa x^{\al\ald}}~~, \qquad
\bar{v}^\pm_i~ = ~{\lm}^\al_\pm\,
\frac{\pa}{\pa{\th}^{\al i}} \qquad
\ee

\subabschnitt{\bf 3.4 BRST formulation in Supertwistor space:}\noindent
 Here we treat the twistorial parameter ${\lambda}^{\al}$ as 
 a constant complex parameter unlike the usual twistor string
theory where the super-twistor space is the target space and also
constrained to be a superconformal theory. In this sense ${\lambda}^{\al}$
can be treated as a zero mode. (However we can always treat 
${\lambda}^{\al}$ and its canonical conjugate  as a conformal theory
with additional ghosts for conformal invariance.)
As mentioned in section 2. $G^{\al \ddz}$ has spin one and is treated as
the BRST current for the left movers and similarly ${\hat G}^{\al \ddz}$
for the right movers. So 
\bee
Q^{\al}_{BRS}\ = \ {\int} dz ~ G^{\al \ddz}(z)~~,\qquad
 {\hat Q}^{\al}_{BRS} ~= ~ {\int}d{\bar z}~~
{\hat G}^{\al \ddz}({\bar z}).
\ee
For the closed string theory let us define
\bee
 Q~=~ {\lm}_{\al}\left(Q^{\al}_{BRS}~ + ~ {\hat Q}^{\al}_{BRS}\right)
\ee
which gives
\bee
 Q~ = 
~{\lm}_{\al}\left[\left(\psi^{\ald\ddz}~+~{\hat{\psi}}^{\ald\ddz}\right)
 \frac{\pa}{\pa x^{\al \ald}} ~+ ~\left( c^i~+ ~ {\hat c}^i\right)
\frac{\pa}{\pa {\th}^{i \al}}\right]
\ee
The BRST operator is equivalent to the $(0,1)$ vector fields as described
in eq.(25). If we assume that the string field also depends on the
twistor variable $\lm$, then inorder that the gauge invariance in the
entire twistor space is maintained we have to add to the {\bf Q} another 
antiholomorphic differential $Q_{\lmb}~=~ d{\lmb}\frac{\pa}{\pa {\lmb}}$.
One can easily show that $Q^2~=~0$, $Q_{\lmb}^2 ~=~0$ and also the  anti-
commutator of $\{Q, ~Q_\lmb\}~=~0$ due to the Berezinian nature of
$\psi$, $\th$ and $\lmb$. 
The desired vertex operator in this BRST cohomology is
\bee
 V~ = ~ 
{\lm}^{\al}\left[\left(\psi^{\ald\ddz}~+~{\hat{\psi}}^{\ald\ddz}\right)
H_{\al \ald}~ + ~\left( c^i~+ ~ {\hat c}^i\right)H_{\al i}\right]~ 
+~d{\lmb}H_{\lmb}.
\ee
From eq.(9) and eq.(11) we show that
 $\psi^{\ald\ddz}$ has conformal spin 
zero and 
charge one and also for $c^i$'s.
(Henceforth we will use $\psi^{\ald\ddz}$
as $\psi^{\ald}$ for convenience).
The vacuum expectation of these zero modes are
\bee
\left<\psi^{\ald}\psi^{\btd}{\Pi}_{i = 1..\CN} c^i_0\right> ~ =~ 
{\eps}^{\ald \btd}
\ee
and also similarly for right moving hatted variables.
Let us define ${\th}^{\al i }~= c_0^1~{\th}^{\al i}$ just to 
make
${\th}^{\al i}$ as fermionic coordinates of charge one.( Here any one of 
the zero modes of $c^i$ will do. We can work on either
of the coordinate patches say on ${\cal U}_{-}$ and
drop the - sign from ${\lm}^{\al}_{-}$).
 For the open  string and $\CN$ = 4, we 
have
\bea 
H_{\al \ald}~ &=& ~{\CA}_{\al \ald}~+~ \th^j_\al ~
{\bar \xi}_{\ald j} +
\th^j_{\al} ~{\th}^{\bt k} ~\pa_{\bt \ald} ~\phi^{lm} ~{\eps}_{jklm}
\nonumber\\[2ex]
&&+ {\th}_{\al}^j~{\th}^{\bt k}~ {\th}^{\gm l}~
\pa_{\bt \ald}~ \xi^m_{\gm} ~
\eps_{jklm}
+(\th^4)^{\bt \gm \de}_{\al} ~\pa_{\bt \ald} ~G_{\gm \de},
\\[2ex]
H_{\al i}~~&=& ~ (\th^j_\al~ \phi^{kl} +\th_\al^j~ \th^k_\bt~ \xi^{\bt l} +
\th^j_\al~ \th^k_\bt~\th^l_\de~ G^{\bt \de})~\eps_{ijkl},
\eea
and
\bea
H_{\lmb} & =& \Bigg(\h ~\nu^2~
 \lm^\al ~\lm^\bt ~\th^i_\bt ~\th^j_\al ~\phi^{kl} +
  \frac{\nu^3}{3!} ~{\lm}^{\al}~{\lm}^{\bt}~{\lm}^{\gm}~ \th_\al^i ~
\th^j_\bt ~\th^k_{\gm}~{\tilde \lm}_{\si}~  \xi^{\si l}~ 
\nn\\[2ex] 
&&+  \frac{\nu^4}{4!}~{\lm}^{\al}~{\lm}^{\bt}~{\lm}^{\gm}~{\lm}^{\de}~
\th^i_\al~ \th^j_\bt~\th^k_\gm~ \th^l_\de ~
 {\tilde \lm}_{\si}~{\tilde \lm}_{\ka}~G^{\si \ka}\Bigg)~\eps_{ijkl}
\eea
Witten's OSFT action is given as
\bee
  S~=~\int Tr \left< V Q V ~+~ {\frac{2}{3}} V^3\right>.
\ee 
where $\int$ denotes integration over all the zero
modes namely $\int d\lm d^4x d^4\th $ and Tr is
 the Trace over the Chan-Paton factors.
Using vacuum expectation values (eq.(30)) and
integrating over the fermionic variables and $d^2\lm$
one gets
\bea
 S~&=&~ \int d^4x\Bigg[ G^{\al \bt}\Big({\pa}_{\al \ald}
{\CA}^{\ald}_{\bt}~+~ {\pa}_{\bt \ald}
{\CA}^{\ald}_{\al}~+~\lb {\CA}_{\al \ald}
, {\CA}^{\ald}_{\bt}\rb \Big)
\nonumber\\[2ex]
&&~+~ {\xi}^{\al j}{\nb}_{\al \ald}
{{\bar \xi}^{\ald}}_ j
~+~{\nb}_{\al \ald}\phi^{ij}{\nb}^{\al \ald}\phi^{kl}
{\eps}_{ijkl}\Bigg],
\eea
where ${\nb}_{\al \ald}~=~ {\pa}_{\al \ald}~+
~\CA_{\al \ald}$. This is the  $\CN$ = 4 self-dual super Yang-Mills action.

\subabschnitt{\bf 3.5 Closed String field theory:}\noindent
 The usual closed string field theory action \cite{HIKKO} is given as
\bee
S = {1\over 2}~V\cdot Q V  + {g\over 3}~V\cdot V \star V \,,
\ee
where $\cdot$ and $\star$ are explained in the appendix.
Also invariant under the BRS transformation
\bee
\delta_{\rm B} V =  Q V + g V \star V .
\ee
Conventional bosonic closed string theory is quite subtle due
to the gost number conservation. However
our case here is further complicated   due to the unusual
closed string  vacuum which we chose  has charge - 20. So gauge 
invariance
is unusual which we will discuss in the sequel.

 For convenience let us rescale each $\th$ variable with
the zero modes of the $c_0^i$. Let us define
\bee
\th^i~ =~ c^1_0~~ {\hat c}^1_0~~ {\lm}_{\al}~~{\th}{\al i}
\ee
where ${\lm}$ is chosen on either of the cordinate patches of
the twistor space. Now each $\th^i$ has charge +2.
To avoid clutter let us define
\bea
\theta^{i}~&=&~\theta^{i}~,\qquad \qquad
\theta^{ij}~=~\frac{1}{2!}\theta^{i}\theta^{j}~,\qquad\qquad
\theta^{ijk}~=~\frac{1}{3!}\theta^{i}\theta^{j}\theta^{k}~,\nn\\[2ex]
\theta^{ijkl}~&=&~\frac{1}{4!}\theta^{i}\theta^{j}\theta^{k}\theta^{l}
~,\qquad \qquad
\theta_{ijk}~=~\frac{1}{5!}\epsilon_{ijklmnop}
\theta^{l}\theta^{m}\theta^{n}\theta^{0}\theta^{p}~, \nn\\[2ex]
\theta_{ij}~&=&~\frac{1}{6!}\epsilon_{ijklmnop}\theta^{k}
\theta^{l}\theta^{m}\theta^{n}\theta^{o}\theta^{p}~,\qquad \qquad
\theta_{i}~=~\frac{1}{7!}\epsilon_{ijklmnop}
\theta^{j}\theta^{k}\theta^{l}\theta^{m}\theta^{n}
\theta^{o}\theta^{p}~, \nn\\[2ex]
\theta^{8}~&=&~\frac{1}{8!}\epsilon_{ijklmnop}
\theta^{i}\theta^{j}\theta^{k}\theta^{l}
\theta^{m}\theta^{n}\theta^{o}\theta^{p}~.\nn 
\eea
The  $2^{\CN}$ component fields will appear in the expansion 
of $H_{\al \ald}$, $H_{\al i}$
 and $H_{\lmb}$ in terms ${\th}^i$ s.
\bea
 H_{\al \ald}& = &  h_{\al \ald}~+~ \th^i~{\tilde \lm}_{\al}~
{\xi}_{\ald i} + \th^{ij}~ {A_{\al \ald}}_{ij} 
+ {\th}^{ijk}~{\tilde \lm}_{\al}~{\chi}_{\ald ijk} +
{\th}^{ijkl}~\pa_{\al \ald}~ \phi_{ijkl}
\nn \\[2ex]
&&+{\th}_{ijk}~{\tilde \lm}_{\bt}~\pa_{\al \ald}~{{\bar \chi}_{\bt}}^{ijk}
+  {\th}_{ij}~{\tilde \lm}_{\al}~{\tilde \lm}_{\bt}
\pa_{\gm \ald}~G^{\bt~ \gm~ ij}
+ {\th}_i~{\tilde \lm}_{\bt}~\pa_{\al \ald}~ {\bar \xi}^{\bt}_i 
\nn \\[2ex]
&&+{\th}^8~ {\tilde \lm}_{\al}~{\tilde \lm}^{\bt}~{\tilde \lm}^{\gm}
{\tilde \lm}^{\de}
  \pa_{\bt \ald}~\omb_{\gm \de}
\eea
Similarly 
\bea
H_{\lmb} & =& 
{\th}^{ijkl}~{\phi}_{ijkl}~+~{\th}_{ijk}~{\tilde \lm}^{\bt}
~{{\bar \chi}_{\bt}}^{ijk}+
{\th}_{ij}{\tilde \lm}_{\al}{\tilde \lm}_{\bt}G^{\al \bt}
\nn \\[2ex]
&&~+~{\th}_i~ {\tilde \lm}_{\al}~ {\bar \xi}^{\al i}
~+~ {\th}^8~ {\tilde \lm}^{\al}~{\tilde \lm}^{\bt}~{\omb}_{\al \bt}
\eea

 Let us first fix  the 
Kinetic energy part of the string field theory action.
As explained in the Appendix the dot product
for the conventional bosonic string fields satisfies
\bee
\Psi \cdot \Phi = \Big< R(1,2)~|~ b_0^{- 2} ~|~ \Psi_2 \Big>~|\Phi_1>
\ee
 due to that the kinetic part of the action ${1\over 2}V\cdot Q V$
is defined as $\int \left<  V ~(c_0 ~ -{\hat c}_0)~ Q V\right> $. When one 
imposes the linear gauge invariance $\de V~~=~~ Q \Lambda$ this will be 
invariant only when $(b_0~- ~{\hat b}_0)V = (b_0~- ~{\hat b}_0)
~\Lambda =~0$. This condition is imposed as the gauge condition on
string field.
In this  case for the vacuum charge to be maintained we prescribe
( as has done in section 3.1,)
\bee
\int \left< V ~ \left(J^{\ddz\ddz}_0 ~+~ {\hat 
J}^{\ddz\ddz}_0\right)~Q~ V \right>\,.
\ee 
Demanding the gauge invariance $\de V~~=~~ Q \Lambda$ we can verify that
the gauge condition happens to be either 
$\left(J^{\ddo \ddo}~-~ {\hat J}^{\ddo \ddo}\right)QV = 0$ or 
in otherwords 
\bee
\left(G^{\al \ddo}~-~ {\hat G}^{\al \ddo}\right)V = 0
\ee
The $G^{\al \ddo}$ are weight 2 operators which is equivalent to
$b_0$. 
$H_{\lmb}$ part of the field does nothave $\psi^{\ald}$ part
and contains all the auxiliary fields. After contractions of 
$\psi^{\ald}$
we get 
\bee
S_{K.E}~ =~ \int d^4x~ d^8{\th}d^2{\lm}\left<
   {\lm}^{\al}~{\lm}^{\bt}{\eps}^{\ald \btd}~ H_{\lmb}
\left({\pa}_{\al \ald}H_{\bt \btd}\right)\right>
\ee
The fermionic integration will give the Kinetic energy part
of all the fields of positive helicity. For the bosonic truncation
let us take only ${\th}^8$ term of $H_{\lmb}$ which gives
\bee
\int d^4x~{\eps}^{\ald \btd} ~{\omb}^{\al \bt} {\pa}_{\al \ald}
~h_{\bt\ btd}
\ee
 The interaction part of the action is ${g\over 3}V\cdot V \star V$
where $\star $ operation and its combinotorics are defined in the 
Appendix. As mentioned earlier three string combinotorics are
$(A(BC)) = ((BA)C) +((AC)B)$. Since $H_{\al \ald}$ is fermionic
 and $H_{\lmb}$ is bosonic we have the interaction term
which  will be glued are in the form
be $\lb H_{\al \ald}H_{\bt \btd}~-~ H_{\bt \btd}H_{\al \ald}\rb
 \cdot H_{\lmb} $. 
Demanding the gauge condition
\bee
  \left(G^{\al \ddo}~-~ {\hat G}^{\al \ddo}\right)
 \Big[ H_{\al \ald}H_{\bt \btd}~-~ H_{\bt \btd}H_{\al \ald}\Big]
  ~= ~0 \,,,
\ee
 shows that the two form $\lb H ,H\rb $ is a constant and hence has a 
poisson
structure. More explicitly this will be satisfied if the
antisymmetric bracket will be taken with each form operated by
its conjugate $\left(G^{\al \ddz}~-~ {\hat G}^{\al \ddz}\right)$.
After performing the vacuum contractions of $\psi$ we get
\bee
 S_{int}~ = ~ \int d^4x~ d^8{\th} ~ d^2 {\lm}
{\lm}^{\al}~{\lm}^{\bt}~{\lm}^{\gm }~{\lm}^{\de}
{\eps}^{\ald \gmd}{\eps}^{\btd \ded}H_{\lmb}\left({\pa}_{\al \ald}
H_{\bt \btd}{\pa}_{\gm \gmd}H_{\de \ded}\right)
\ee
So far we have not taken the gauge part of the fermionic cordinate.
Similarly we shall expand $H_{\al i}$ as in the previous case for the 
$\CN$ =4 super Yang-Mills. Going through the same excercise
we shall get interaction term in the super space as
\bee
S_{int}~ = ~ \int d^4x~ d^8{\th} ~ d^2 {\lm}~
{\lm}^{\al}~{\lm}^{\bt}~{\lm}^{\gm }~{\lm}^{\de}
 W^{A B}~W^{C D}~H_{\lmb}~\left({\pa}_{\al A}
H_{\bt C}~{\pa}_{\gm B}~H_{\de D}\right)
\ee
where $W^{A~B}$ is the metric in the super-space as $\{ {\eps}^{\ald 
\btd}~,~ {\de}_{i j}\}$. This is  the main  result which resembles the 
Plebanski 
equation in the super-space. One can make the bosonic truncation after
taking only the ${\th}^8$ component of the $H_{\lmb}$ and get the usual
Plebanski equation. One also get the Prepotential equation of Karanas
and Ketov \cite{KK} once one identifies $H_{\al ald}~ =~ {\pa}_{\al \ald}
\Phi $ where $\Phi$ is the prepotential. 

\subabschnitt{\bf 4. The Conclusion:}\noindent
 We have given here a closed string field theory formulation of
$\CN$ = 8 self-dual supergravity. This is completely off-shell and
Lorentz covariant. The field theory limit gives the correct action
which was earlier formulated by Karanas and Ketov\cite{KK} and recently
by Mason and Wolf \cite{MW} in the super twistor space. Unlike the
 conventional twistor string formulation of recent times 
\cite{WITT,BERK1} and \cite{ BEWT,NA}
where twistor space is the target space we treat here the twistor 
variable as a complex parameter of mixing the world sheet super 
conformal generators. We will calculate the N point MHV amplitudes
using  the Witten and Nair's \cite{WITT,NA} principle which  enables 
us to calculate the higher point  function in the twistor space 
although without the twistor space more than 3 point function 
vanishes.

\abschnitt{\bf Appendix}
 In this appendix we present very basic properties of Closed string field 
theories.For any generic  string functionals $\Phi_i$ ($i=1,2,3$),
 the dot and the
star products have  the following properties:
\bea
\Phi_1\cdot\Phi_2 &=& (-)^{\abs{1}\abs{2}}\Phi_2\cdot\Phi_1 ,
 \\[2ex]
\Phi_1\star\Phi_2&=& -(-)^{\abs{1}\abs{2}}\,\Phi_2\star\Phi_1 ,
 \\[2ex]
\Phi_1\cdot\left(\Phi_2\star\Phi_3\right)
&=&(-)^{\abs{1}\left(\abs{2}+\abs{3}\right)}
\,\Phi_2\cdot\left(\Phi_3\star\Phi_1\right) ,
\\[2ex]
\Phi_1\cdot Q_{BRS}\Phi_2&=&-(-)^{\abs{1}}\left( Q_{BRS}\Phi_1\right)
\cdot\Phi_2 ,
\eea
where $\abs{i}$ ($i=1,2,3$) is $0$ ($1$) if $\Phi_i$ is Grassmann-
even or odd.

\subabschnitt{\bf A1. The Reflector:}\noindent
Let $\mid \Phi_i>$ is expanded as a basis in a Hilbert space and
let $<\Phi_i\mid$ be its dual then
 $\left<\Phi_i\mid \Phi^j\right> = \de_{ij}$.
Thus the dot product is defined as 
\bee
\Psi \cdot \Phi = \left< R(1,2)\mid \Psi_2 >\mid\Phi_1\right>
\ee
The reflector maps the state $\mid O>$ to its BPZ conjugate state.
More explicitly  if two string states are described by two punctured
speheres with  a uniformizing cordinate say z. The local coordinate 
$z_1=z=0$ and another with local cordinate $z_2 = \frac{1}{z}=0$
So 
\bee
\left< R(1,2)\mid \Psi_i (1)>\mid\Phi_j (2)\right> =
 \left<\Psi_i(z_2=0) \Phi_j (z_1=0)\right> ~ = ~~ G_{ij} 
\ee
This will be  non-vaishing of  only when the total ghost number of
the matrix element vanishes. The closed string vacuum has ghost
number - 6. Since any physical bosonic closed string has ghost number
2, inorder that the dot product will be non-vanishing one defines that
\bee
\Psi \cdot \Phi = \left< R(1,2) b_0^{-~ 2}\mid \Psi_2 >\mid\Phi_1\right>
\ee
where $b_0^- = b_0 -{\hat b}_0$.

\subabschnitt{\bf A2. The $\star$ Product:}\noindent
The  $\star$ product is defined as mapping from two string fields to
one string field such as 
\bee 
\mid \Phi> \mid \Psi> = \mid \Phi\star \Psi>
\ee
For the three point vertex in the language of conformal field
theory \cite{LPP}, which is written as three conformal mapping
of $\{h_r \mid r= 1,2,3\}$ from a unit disk to sphere with
cordinates z as
\bee
\left<{\mathcal V}_{123}\mid \mid\phi_1>\mid\phi_2>
\mid phi_3\right> = \left< h_1(\phi_1)(z_1)
h_2(\phi_2)(z_2)h_3(\phi_3)(z_3)\right>_{S_2}
\ee
Each map $h_r$ is a conformal transformation of the vertex operator
at the origin on the disk to one on the sphere.

 \end{document}